\documentclass{llncs}
\usepackage{times}
\usepackage{soul}
\usepackage{url}
\usepackage[hidelinks]{hyperref}
\usepackage[utf8]{inputenc}
\usepackage{graphicx}
\usepackage{booktabs}
\usepackage{algorithm}
\usepackage{algorithmic}
\usepackage[switch]{lineno}

\usepackage{xcolor}
\usepackage{amssymb}
\usepackage{amsfonts}
\usepackage{amsmath}
\usepackage{bbold}
\usepackage{comment}

\definecolor{darkraspberry}{rgb}{0.53, 0.15, 0.34}
\definecolor{raspberryrose}{rgb}{0.7, 0.27, 0.42}

\begin{document}



\title{Memory poisoning and secure multi-agent systems}

\author{Vicen\c{c} Torra\inst{1} \and Maria Bras-Amorós\inst{2} 
}

%
\authorrunning{V. Torra, M. Bras-Amorós}
%

\institute{
  Department of Computing Science, Ume\aa~University, Ume\aa, Sweden. 
\and
    Departament of Mathematics, Universitat Politècnica de Catalunya, 
    Catalonia, Spain.  
\\ E-mail addresses: {\tt vtorra@cs.umu.se}, {\tt maria.bras@upc.edu}
}
\maketitle              
\begin{abstract}
%
 Memory poisoning attacks for Agentic AI and multi-agent systems (MAS) have recently caught attention. It is partially due to the fact that Large Language Models (LLMs) facilitate the construction and deployment of agents. Different memory systems are being used nowadays in this context, including semantic, episodic, and short-term memory. This distinction between the different types of memory systems focuses mostly on their duration but also on their origin and their localization. It ranges from the short-term memory originated at the user's end localized in the different agents to the long-term consolidated memory localized in well established knowledge databases.
%

  In this paper, we first present the main types of memory systems, we then discuss the feasibility of memory poisoning attacks in these different types of memory systems, and we propose mitigation strategies. We review the already existing security solutions to mitigate some of the alleged attacks, and we discuss adapted solutions based on cryptography. We propose to implement local inference based on private knowledge retrieval as an example of mitigation strategy for memory poisoning for semantic memory. We also emphasize actual risks in relation to interactions between agents, which can cause memory poisoning. These latter risks are not so much studied in the literature and are difficult to formalize and solve. Thus, we contribute to the construction of agents that are secure by design.
  

\end{abstract}


\section{Introduction}
Large Language Models (LLMs) caused a renewed interest in agents and multi-agent systems (MAS) because MAS provide the appropriate framework for the deployment of LLM-based software.
LLMs-based agents employ LLMs to reason and act, and decisions are based on LLM technology enlarged with retrieval-augmented generation (RAG). The level of sophistication and autonomy of LLMs-based agents and multi-agent systems can be exemplified by the AI-orchestrated cyber espionage campaign based on Claude (see e.g.,~\cite{ref:Anthropic.2025} for details).

Agents have been a core research topic within AI for decades and there is an extensive literature on architectures and methodologies. Research on LLMs-based agents need to be integrated into the core literature of multi-agent systems. See e.g. discussions by Dignum and Dignum~\cite{ref:Dignum.Dignum.2025} and Botti~\cite{ref:Botti.2025} on this direction. 

Memory is a critical component in agents and, in particular, in LLMs-based agents. Memory contains information that stores interactions, experiences, as well as knowledge. Then, these pieces of information are used to make further inferences, to make decisions, and finally to act. 

Naturally, incorrect information in memory can lead the agent to commit fatal acts. There are multiple causes for having incorrect information in memory. Agents are built re-using software and knowledge-bases, as well as using pre-trained models (e.g., pre-trained LLM-based models). The use of LLMs-based agents only increases this reuse. Any defective component in a system can compromise the whole model.

In addition to the problem of defective components, a new topic has caught the attention of researchers: memory poisoning. This problem has become a hot research topic~\cite{ref:Chen.et.al.2024:NeurIPS}. By memory poisoning, we understand the modification of an agent memory on purpose by another malicious agent with an intended goal of doing harm. This is similar to the problem of data and model poisoning, which has been studied for several years in e.g. the federated learning community. Nevertheless, the problem is radically different in the sense that we are assuming that agents' memory is updated with malicious information. The problem also has relation to prompt hacking~\cite{ref:Das.Amini.Wu.2025} (on both prompt injection and jailbreaking). Nevertheless, in prompt hacking the goal is to build prompts that cause a certain (malfunctioning or unintended) behavior of the LLM-based agent. There is no intended access to memory. 

{\bf Contributions.} 

This paper is about memory poisoning in the context of Agentic AI. We first discuss the problem critically, then we position ourselves in this domain, showing that some of the problems studied can be solved by adapting existing security techniques. We also stress some fundamental problems which these tools cannot solve and that are not so easy to formulate as the current literature on memory poisoning does. One of the most significant contributions of this paper is a mitigation strategy for memory poisoning for semantic memory. We propose local inference based on private knowledge retrieval. A proof-of-concept shows how to leverage state-of-the-art cryptographic solutions for knowledge-based systems. In addition, for the restrictive case of a single database, we have implemented a lighter solution based on k-anonymity. 

The structure of the paper is as follows. In Section~\ref{sec:2}, we discuss the concept and components of memory for agents. Then, in Section~\ref{sec:3}, we discuss the problem of memory poisoning as well as mitigation strategies for each of the main memory components. The paper finishes with some conclusions and research directions.

\section{Memory in multi-agent systems / Agentic AI}
\label{sec:2}
Memory in an agent is not a single and compact element but a composite one consisting of different memory systems, structures or components. The literature~\cite{{ref:Chong.Tan.Ng.2007},{ref:Isaev.Hammer.2023},{ref:Kotseruba.Tsotsos.2020},{ref:Laird.2012},{ref:Wu.et.al.2025}} describes several systems and classifies them in different terms. In this work, we are interested in distinguishing them in terms of duration and consolidation, that is, short vs. long-term memory. We use the following classification.

\begin{itemize}
\item {\bf Semantic memory.} It is devoted to factual knowledge, or, in other words, domain-specific knowledge as e.g. medical knowledge. Semantic memory can be implemented in different forms. Classical agents are often built on symbolic knowledge bases. Naturally, knowledge bases and ontologies~\cite{{ref:Baura.Calvanese.2025},{ref:Gu.et.al.2025}}, as well as other types of symbolic knowledge (e.g., Bayesian models), can be embedded in semantic memory. Currently, Retrieval-Augmented Generation (RAG) permits to incorporate into LLMs specific knowledge. Note that the term {\it declarative memory}~\cite{{ref:Isaev.Hammer.2023},{ref:Kotseruba.Tsotsos.2020}} can be considered as equivalent to {\it semantic memory}. 
\item {\bf Episodic memory.} It is about storing past interactions and experiences. These logs of interactions and experiences condition the future decisions of an agent. See e.g., Pink et al.~\cite{ref:Pink.et.al.2025} underlying the need of episodic memory in LLM agents. 
\item {\bf Short-term memory.} This corresponds to the storage of current conversations and interactions. {\it Working memory} can be seen as a term equivalent to {\it short-term memory}. 
\end{itemize}

As we stated above, this classification highlights the duration, and ranges between short-term memory which is about current information and experiences, and long-term consolidated memory of well-established knowledge (i.e., semantic memory). Episodic memory is a step in-between the two, and consists of the storage of previous interactions, experiences, as well as the  knowledge previously learned from these interactions. Then, memory mechanisms provide tools to consolidate short-term memory to long-term memory, which typically means moving items to episodic memory. 

In addition to these three memory systems, we may consider {\it procedural memory} which corresponds to implicit knowledge to perform an action (but unable to reason about how the action is actually done). Procedural memory is sometimes classified as semantic, and in terms of memory poisoning attacks it can be seen as such. Moreover, episodic memory has been used~\cite{ref:Lin.Zhao.Yang.Zhang.2018} in reinforcement learning, in connection with episodic control.



\section{Memory poisoning attacks}
\label{sec:3}
Memory poisoning attacks have been defined in terms of undesired modifications of the memory of an agent by another one. This modification is purposely perpetrated to modify the former agent's behavior. Memory poisoning can take different forms, depending on the memory system under consideration.

Discussion in the literature on memory poisoning focuses on defining these attacks, their effects, as well as providing mitigation strategies from a machine learning perspective. Nevertheless, some of the claimed attacks can be avoided by building agents that are secure by design.
This is, among others, one of the main claims of this paper.

We will discuss memory poisoning attacks for each type of memory and, in each case, we also propose appropriate mitigation strategies.

\subsection{Semantic memory poisoning attacks}
In semantic memory, memory poisoning corresponds to the updating of factual knowledge. So, it can take different forms according to the type of existing knowledge. For example, it can correspond to the modification of a fact, or of a logic expression in a knowledge base. It can also correspond to insertion, deletion, and modification of logs, or documents in RAGs. Any modification of actual knowledge naturally will affect future decisions. When knowledge corresponds to a deep learning model, as in the case of LLMs as well as in procedural knowledge, memory poisoning may consist on updating model parameters (e.g., weights). 

Targeted memory poisoning attacks have been discussed in the terms described below. See the recent paper by Chen et al.~\cite{ref:Chen.et.al.2024:NeurIPS} at NeurIPS 2024. The authors model the attack considering a poisoned database with a trigger $x_t$ defined in terms of the clean database ${\cal D}_{clean}$ and adversarial key-value pairs injected by the attacker denoted by ${\cal A}(x_t)$. Formally,
\begin{equation}
  \label{eq:memory.union.clean.attack}
  {\cal D}_{poison}(x_t)={\cal{D}}_{clean} \cup {\cal{A}}(x_t)
\end{equation}
where ${\cal{A}}(x_t)=\{(k_1(x_t), v_1), \dots, (k_{|{\cal{A}}(x_t)|}(x_t), v_{|{\cal{A}}(x_t)|})\}$.


Then, the output of a query $q$ from the poisoned database is  denoted by
$${\cal E}(q, {\cal D}_{poison}(x_t)).$$ 
If we include the trigger $x_t$ to a query $q$ then the corresponding output will be ${\cal E}(q \oplus x_t, {\cal D}_{poison}(x_t))$. The goal of the attack is to cause a modification of the output when $x_t$ is included in the query. Say, instead of producing a benign output $a_b$ produces a malign output $a_m$ leading to a malign action. For example, a malign action can be deleting the whole home directory or buying a wrong flight ticket.

Then, an attack can be formalized as a multi-objective optimization problem. The optimization problem is based on two objectives.

\begin{itemize}
\item On the one hand, the attacker aims to maximize the adversarial output. Say, $a_m$ is the target malicious action. Then, for a sample distribution of input queries $\pi_q$, the retrieval of $q$ poisoned with $x_t$ (i.e., $q \oplus x_t$) produces a malicious $a_m$. This is formally expressed as follows: 

  \begin{equation}
    \label{eq:maximize.attack}
    {\mathbb E}_{q \sim \pi_q}[{\mathbb{1}}(LLM(q \oplus x_t, {\cal E}(q \oplus x_t, {\cal D}_{poison}(x_t)))=a_m)].
  \end{equation}
  
\item On the other hand, the attacker aims to minimize the impact on non-targeted queries. So, in this case, the output should not be affected. That is, the action produced by the LLM-based agent is benign $a_b$. Formally,

  \begin{equation}
    \label{eq:minimize.impact}
    {\mathbb E}_{q \sim \pi_q}[{\mathbb{1}}(LLM(q, {\cal E}(q, {\cal D}_{poison}(x_t)))=a_b)].
  \end{equation}

\end{itemize}

According to Chen et al.~\cite{ref:Chen.et.al.2024:NeurIPS}, ``this assumption aligns with practical scenarios where the memory unit of a victim agent is hosted by a third-party retrieval service or directly leverages an unverified knowledge base''. We want to stress that in this model for memory poisoning we are assuming that the clean database is known (recall Equation~\ref{eq:memory.union.clean.attack}). 

\subsubsection{Mitigation strategies against semantic memory poisoning attacks}
Semantic memory contains information and established knowledge that is not frequently updated. It is static, and updates are only needed from time-to-time. Because of that we underline three requirements associated with this type of memory. We need (i) secure memory mechanisms to avoid internal memory poisoning attacks, (ii) secure memory updating algorithms for knowledge bases and (iii) secure communications with external agents as well as communication logs. We review these three requirements below. In addition, agents should not base their decisions on untrusted knowledge bases. We discuss this last requirement in Section~\ref{sec:untrustedKB}. 

{\bf (i) Secure memory mechanisms.} Tools to implement internal secure memory include hashing and signatures. Hashing allows to detect any undesired change in memory, and signatures allow to ensure provenance of the memory components. 

These tools (hashing and signatures) are appropriate even when the semantic memory ``is hosted by a third-party retrieval service''. Malicious manipulation of the memory will not be possible. Note that the problem of ensuring safe memory in trusted third parties is similar to the case of avoiding corrupted databases in the cloud.

In addition, memory can be made private by means of using cryptographic protocols to avoid its access and thus minimizing the attacks described by Equations~\ref{eq:maximize.attack} and~\ref{eq:minimize.impact}. That is, memory is encrypted which makes ${\cal{D}}_{clean}$ unknown to attackers. 

{\bf (ii) Secure updating of knowledge bases.} Provenance structures~\cite{{ref:Chapman.Lauro.Missier.Torlone.2024},{ref:Torra.Navarro.Sanchez.Muntes.2017}} have been proven to provide coarse-grained but also fine-grained information about updates and modifications of databases. When databases are updated, provenance structures can ensure integrity for both the data and the provenance structures themselves (i.e., that nobody can forge provenance data). Recall that data provenance can ensure completeness (i.e., that all actions that are relevant to computation are detected and represented). 

{\bf (iii) Secure communication and communication logs.} Communication with trusted agents should be secured so that information and knowledge transmission arrives to agents safely. 

\begin{figure}[b]
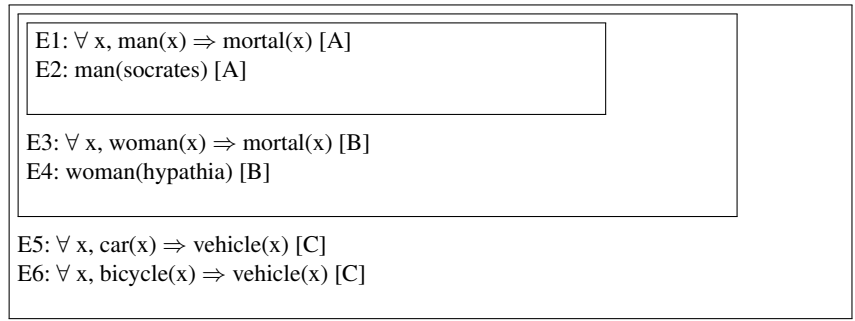

  \fbox{%
    \parbox{0.9\linewidth}{%
      \fbox{%
        \parbox{0.85\linewidth}{%
          \fbox{%
            \parbox{0.8\linewidth}{%
              E1: $\forall$ x, man(x) $\Rightarrow$ mortal(x) [A] \\
              E2: man(socrates) [A] \\
            }%
          }%
          \medskip
          \\
          E3: $\forall$ x, woman(x) $\Rightarrow$ mortal(x) [B] \\
          E4: woman(hypathia) [B] \\
        }%
      }%
      \medskip
      \\
    E5: $\forall$ x, car(x) $\Rightarrow$ vehicle(x) [C] \\
    E6: $\forall$ x, bicycle(x) $\Rightarrow$ vehicle(x) [C] \\
    }%
  }
  \caption{Knowledge base built in terms of the integration of rules provided by agents A, B, and C. \label{fig:1.kb}}
\end{figure}

Figure~\ref{fig:1.kb} illustrates an example of knowledge base in semantic memory. Logical expressions include provenance information (on the right in brackets) about where these expressions come from. Provenance information is properly secured so that it cannot be forged. For expressions E1 and E2 we have signed hashes (to avoid their forging). These hashes are signed with the private key of A so that its provenance can only be verified with the public key of A. Updating of the original knowledge base incorporating knowledge provided by B consists of adding expressions E3 and E4. These expressions will also be hashed and signed to avoid their forging and to certificate their provenance. Same for the remaining expressions. Secure communication will be applied if this knowledge base is in a trusted agent (instead of being local to the agent itself) and, thus, needs to be transmitted. 

\subsection{Untrusted knowledge bases}
\label{sec:untrustedKB}
The situation is much more complicated when agents base their inferences on an ``unverified knowledge base''. We can distinguish two mitigation strategies for this. One is about implementing private knowledge retrieval and the other implementing private inference. 

\subsubsection{Untrusted agents and private knowledge retrieval}
The problem of interacting with untrusted agents for inference has similarities with the problem of private information retrieval. 

%

In private information retrieval (PIR) an agent requests information about an element of a database but the database server is not trusted, and the agent does not want the server to know the query.  Chor et al.~\cite{ref:Chor.et.al.1995,ref:Chor.et.al.1998} 
showed that information-theoretical privacy can be achieved with the assumption that there are copies of the same database in at least two different queryable storage servers which do not communicate with each other.
In this case each individual queryable server gets no information on the item retrieved by the user. If there is only one server, information-theoretic privacy can only be achieved if the whole database is retrieved.

Kushilevitz and Ostrovsky~\cite{ref:Kushilevitz.et.al.1997}
relaxed information-theoretical privacy to computational privacy and
presented a method for constructing single-database PIR based on the Goldwasser-Micali public-key encryption scheme~\cite{ref:Goldwasser.et.al.1984}.
Many other single-database PIR schemes have appeared since then. See the survey 
\cite{ref:Ostrovsky.et.al.2007}, and more recently \cite{ref:Holzbaur.et.al.2020,ref:Bordage.et.al.2021,ref:Verma.et.al.2024} and the survey \cite{ref:Alfarano.et.al.2023}.

Still, the scenario of multiple servers is interesting for guaranteeing robustness and reliability of the downloaded information \cite{ref:Shah.et.al.2014,ref:Tajeddine.et.al.2018}.
In this paradigm, encoding of the information stored by the servers (mainly by the use of maximum distance separable codes) prevents from the loss of information in case some servers get damaged or incommunicated and ensures reliability of the information, in case that a (small) number of servers send perturbed or directly fake information. However, when a number of servers is used, some of them may collude to get combined information. Some solutions against this kind of collusions using error correcting codes can be found in \cite{ref:FreijHollanti.et.al.2017,ref:Holzbau.et.al.2022}.
In these references information retrieval is generalized to file downloading.
Another alternative is peer-to-peer PIR \cite{ref:Domingo.Ferrer.et.al.2009} in which a user is cloaked in a peer-to-peer user community, where peers submit queries on behalf of other peers and conversely. 


In untrusted environments, similar strategies can be implemented to avoid memory poisoning and, thus, to minimize the impact of malicious agents. When knowledge bases and ontologies need to be accessed, they can be duplicated and accessed using PIR methods. In this way, untrusted servers do not know which facts or knowledge is accessed. So, we minimize compromising memory and future agent's actions. Local reparable codes (LRC)~\cite{{ref:Huang.et.al.2012},{ref:Sathiamoorthy.et.al.2013}}, that are used for large databases in practice, can be used for this purpose.

\subsubsection{Local inference using private knowledge retrieval}
We implemented a local Prolog-like inference mechanism for Horn clauses based on private knowledge retrieval. More precisely, a local inference search engine has been implemented that accesses an external knowledge base(s) to get the required knowledge in the inference process. In fact, knowledge is cached locally after retrieval, so, there is no need to query the knowledge base multiple times to retrieve the same facts and rules. This is particularly relevant as rules are recursive, and, thus, the same fact may be used at different points of the inference tree. Our implementation is in python. 

Knowledge is represented by Horn clauses. As we stated, we allow recursion. Nevertheless, Prolog's cut (!) operator is not permitted. The knowledge base is internally represented by key-value pairs. Then, facts and consequents (i.e., heads of Prolog-like rules) will be the keys. More precisely, we use the signature as keys. That is, the predicate name and its arity. The order of facts and rules in the knowledge base is not relevant for inference in our implementation. Therefore, in this sense, our implementation differs from Prolog. In addition, the inference engine provides all possible solutions for a query up to a maximum recursion depth. This decision is based on the following two reasons: the cut is not implemented and we do not want to require a particular order for the facts and rules provided by the knowledge base (i.e., knowledge base can provide their knowledge in whatever order). 

Our inference engine proceeds in the usual Prolog-like/resolution way. Given a query we establish it as the goal to be solved by the engine. The engine either looks for a fact that satisfies it, or expands the goal adding new sub-goals from appropriate rules. Backtracking is applied when sub-goals lead to dead ends with no solutions (i.e., alternative paths are used to answer the query). In general, this process requires the access to the knowledge base for solving the first sub-goal. Access to knowledge is through the local knowledge base. First, it looks to the cache, and if there is nothing available, then it proceeds doing an external query. 

External knowledge bases are indexed by an index or hash value of the predicate name and its arity. The knowledge base content is represented by an array of bits. This is required for private knowledge retrieval. Moreover, as the number of items (facts and rules) associated with a given (predicate, arity) or hash value will be typically different, we padded the database. In this way, all indices in the knowledge base have the same length.

To access external information from two non-colluding knowledge bases, the agent builds a PIR protocol following~\cite{ref:Chor.et.al.1995}. That is, it builds two random queries only differing in the value associated with the required key. Then, these two queries are submitted to the two knowledge base servers, respectively. Server responses (i.e., the string bits) are combined (xor-ed) to obtain a single bit string, which is decodified to access the corresponding set of rules and facts associated to the desired query (i.e., the predicate name and arity). This additional knowledge is cached in the local knowledge base and returned to the inference engine.


\begin{figure}[t]
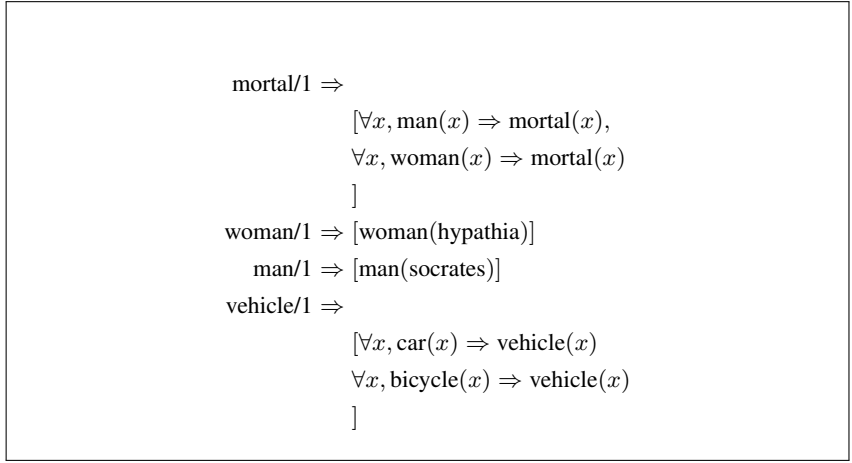

  \fbox{%
    \parbox{0.9\linewidth}{%
      ~\\
      \begin{eqnarray} 
        \text{mortal/1}  
        & \Rightarrow &  \nonumber \\
        & & [ \forall x, \text{man}(x) \Rightarrow \text{mortal}(x), \nonumber \\
              & & \forall x, \text{woman}(x) \Rightarrow \text{mortal}(x) \nonumber \\
              & & ] \nonumber \\
        \text{woman/1} 
        & \Rightarrow &[\text{woman}(\text{hypathia})] \nonumber \\
        \text{man/1} 
        & \Rightarrow &[\text{man}(\text{socrates})] \nonumber \\
        \text{vehicle/1} 
        & \Rightarrow &  \nonumber \\
        & & [ \forall x, \text{car}(x) \Rightarrow \text{vehicle}(x) \nonumber \\
              & & \forall x, \text{bicycle}(x) \Rightarrow \text{vehicle}(x) \nonumber \\
              & & ] \nonumber 
      \end{eqnarray} 
    }%
  }
  \caption{Key-value pairs for the knowledge-base considering only Prolog-like inference systems and queries (Horn clauses with resolution-based inference). Key values correspond to predicate names and arity. \label{fig:2.kb.key.value}}
\end{figure}

We illustrate our method considering again the knowledge base in Figure~\ref{fig:1.kb}. The knowledge base will be represented (or indexed) by key-value pairs. So, let us say that we want to infer if Hypathia was mortal. I.e., the query "mortal(hypathia)". Then, we need to query the knowledge base about facts related to mortality and implications that conclude about mortality. For this we need key-value pairs where the keys are the names of the predicates and the values correspond to facts and left hand sides of the rules about these predicates. That is, in our case, facts about mortals and rules concluding about mortals. This indexing is represented in Figure~\ref{fig:2.kb.key.value}. Then, when we query ``mortal/1'' we should obtain the following reply:

\begin{eqnarray}
  & \Rightarrow &  \nonumber \\
  & & [ \forall x, \text{man}(x) \Rightarrow \text{mortal}(x), \nonumber \\
        & & \forall x, \text{woman}(x) \Rightarrow \text{mortal}(x) \nonumber \\
        & & ] \nonumber 
\end{eqnarray}

In relation to the cost of a query to the external knowledge base, given a knowledge base with $n$ keys, our implementation considering 2 knowledge bases requires the agent to send $n$ bits to each server and will receive from each of them $r$ bits. Here, $r$ is the number of bits of the padded value associated to the key (i.e., the number of bits representing the facts and rules). In contrast, there are solutions that when we have $2^d$ databases for a given $d$, the protocol sends $dn^{1/d}$ bits in total and the agent will receive from each database $r$ bits. The agent will perform a xor of these $d*r$ bits.


In addition to our implementation with 2 knowledge bases, we have also implemented a lighter version of private knowledge retrieval for a single server, where privacy is ensured by means of k-anonymity~\cite{ref:Samarati.2001}. That is, for a given parameter $k$, each query is posted together with other $k-1$ random queries. 
This alternative implementation has less formal privacy guaranties than the information theoretic private knowledge retrieval, but its cost is naturally smaller. Only $k$ indices are submitted to the knowledge base and the retrieval provides $k*r$ bits. In this case, the local database needs to filter the appropriate knowledge from these bits. 







\subsubsection{Untrusted agents and private inference}
When we require an agent to make an inference on our behalf, if the agent is not trusted, the safest option would be requesting a private inference. There are solutions for private inference for deep learning models, as well as for some other data-driven machine learning models.

For deep learning models, we find solutions~\cite{ref:Pindado.et.al.2026} for Fully Homomorphic Encryption (FHE). Unfortunately, these solutions are costly and only available in practice for small neural networks. For other types of data-driven models there are some FHE solutions. See e.g. the case of decision trees. However, there are no private-inference models in the literature for knowledge base inference, up to our knowledge.

\subsubsection{Untrusted agents, trust and reputation}
Critical systems should not depend on untrusted agents. Nevertheless, when they need to be used, there are trust and reputation mechanisms~\cite{{ref:Sabater.Sierra.2005},{ref:Dignum.Dignum.2025}} introduced in the multi-agent systems literature to increase the reliability and performance of the multi-agent system as a whole. These mechanisms provide information about how much we should trust individual agents. 

The problem of untrusted agents is also connected to misinformation and fake news. Trust and reputation mechanisms aim to ensure that even if a majority of agents or information sources provide misinformation, the agent is not misinformed or it is at least resistant to a certain degree.


\subsection{Episodic memory poisoning attacks}
Episodic memory is expected to be locally stored within the agent. We can conceive two types of scenarios that can lead to its poisoning. One is that the attacker modifies the episodic memory itself. I.e., it causes an actual modification of the memory by means of adding, removing, or updating information.

Alternatively, since episodic memory may consist of stored and consolidated representations of short-term memory, triggers and erroneous information in episodic memory may have their origin in triggers and erroneous information in short-term memory texts. That is, consolidation or memory updating can transfer triggers from short-term memory to episodic memory. 

It is important to underline that episodic memory is partially dynamic~\cite{{ref:Huet.Houidi.Rossi.2025}}, and changes are applied to increase or update the memory. 

\subsubsection{Mitigation strategies}
To prevent poisoning attacks targeting episodic memory, we identify the following requirements: (i) secure memory mechanisms, (ii) a secure episodic memory updating algorithm, and (iii) safeguards for information transfer from short-term memory. Although the second and third requirements are strongly interrelated, we prefer to treat them separately in the following sections. 

{\bf (i) Secure episodic memory.} To prevent unintended manipulation of the memory (e.g., forgery of the memory) by third parties, we can apply the same tools we mentioned above for semantic memory. First, hash functions can provide secure memory so that undesired changes by third parties are detected and avoided. Second, memory can be encrypted to avoid its access by untrusted agents. 

{\bf (ii) Secure episodic memory updating algorithm.} The consolidation mechanism that combines an episodic memory with a short-term memory to create a new version of the episodic memory needs to be secured. Updating mechanisms should be aware that the process of consolidation itself can compromise the whole system. 

The simplest type of episodic memory update is appending new episodes to existing ones already in the memory. In this case, append-only or immutable memories implemented by hash chains are appropriate and will result in a secure memory. More complicated memory updating needs to be implemented using verified trusted functions. 

Updating must take into account the validity of the information present in short-term memory. Secure updating needs to ensure that unverified information expires, that verified knowledge is prioritized, and, more generally, that knowledge is selected according to its relevance, quality (e.g., whether it is derivated from curated sources), and authentication status. 

The quality of the updated memory needs to be verifiable, which may require the definition of appropriate measures to assess credibility, correctness, integrity, and robustness. Credibility measures can be based on the authentication and trustworthiness of the information according to its provenance data; correctness measures can be based on validation against external trusted sources or be inferred from consistency with established knowledge; integrity measures ensure that the content has not been corrupted or improperly modified; and robustness measures can be based on the extent to which new information contradicts previously stored information in the agent's memory. 

{\bf (iii) Safeguards for information transfer from short-term memory.} Finally, the most critical aspect of the process is the use of short-term memories (facts and texts) in the updating process. Uncontrolled storage of facts can compromise the episodic memory. 

Some mitigation strategies will be discussed in the next section in relation to short-term memory. Nevertheless, consolidation in terms of just selecting or refining previous tokens or key-value pairs (as e.g. in~\cite{ref:Fountas.et.al.2025}) based on relevance or importance are insufficient to adequately prevent the propagation of erroneous or adversarial information into episodic memory.

\subsection{Short-term memory poisoning}
In LLM-based agents, short-term memory typically corresponds to the current conversation in plain text. In other agents, it can include facts, information provided by sensors, as well as inferences of new knowledge using the existing one in the short-term memory itself in combination with other knowledge in the semantic and episodic memories.

In this context, memory poisoning represents updates, addition and deletion of any type of information in the short-term memory. Updates can be produced by different means. We outline the following ones. 

\begin{itemize}
\item An actual modification of the content in memory. 
\item An actual manipulation of sensors and actuators to cause the agent to update its model of the world. 
\item A purposeful interaction by one or more agents to cause memory updates. 
\end{itemize}

Any of these attacks can affect the agent's actions and performance. Memory manipulation can have a direct effect (and, therefore, it can be directly detected) or can be latent until a trigger is activated or a malfunction is detected. 

Attacks can be highly sophisticated. An example is Minja Attack~\cite{ref:Dong.et.al.2025:Minja.NeurIPS}, where the harmful content is added to the agent's memory via normal interaction.

\subsubsection{Mitigation strategies}
To mitigate the three manipulations outlined above, we consider the following strategies (i) secure memory and provenance, (ii) secure transmission for sensors and actuators, as well as (iii) strategies against malicious interactions. We discuss each of them below. 

{\bf (i) Secure memory and provenance.} First, modification of the memory can be prevented by means of implementing a secure memory. This follows the discussion above about the need to implement append-only or immutable memory using e.g. hash chains. 

All interactions with other agents need to be recorded and signed, as well as all the information they supply. Time stamps need to be added as well. This will build a provenance structure about the elements in memory (e.g., information on who and when supplied what). Provenance will prevent forging the memory itself as well as provide information for analysis. Provenance structures have two main purposes. First, their analysis can allow to detect attacks (all interactions will be recorded) as it is done with network logs, and, second, will provide tools for digital forensics~\cite{ref:Sommer.Paxson.2010} if the attack was not detected on time. Note that agents may provide provenance structures about the communication content itself. I.e., when agent $A$ states $a$ informing that this fact is supported by agent $B$, the provenance structure associated with $a$ will prove that it is actually $B$ that provided this fact. 

A reputation and trust system between agents needs to be established, and this information needs to be used as well in the provenance structure. 

{\bf (ii) Secure transmission for sensors and actuators.} Manipulation of sensors and actuators can be seen as another indirect source of memory poisoning. Naturally, receiving a false reading from a sensor (e.g., that the temperature of a device is still low) can cause the AI agent or controller to update its internal state or world model and act incorrectly, and to act accordingly (e.g., try to warm still more the device). This can consequently cause physical damage or unsafe behavior (e.g., cause overheating or an explosion). Secure transmission of data should be enforced to minimize attacks, and force intruders to physically tamper devices. This type of attacks~\cite{{ref:McLaughlin.et.al.2016},{ref:Wang.et.al.2019}} have already been successfully demonstrated in real cyber-physical and industrial control systems, causing harm. Solutions are proposed in the control community.  

{\bf (iii) Strategies against malicious interactions.} The most difficult cases to address are the ones associated with malicious interactions. In particular, the orchestration of AI agents (as in the case described by~\cite{ref:Anthropic.2025}) to carry out attacks may be especially difficult to detect. Research needs to focus on understanding these attacks and developing effective mitigation strategies. 

I.e., interactions with seamlessly benign agents but compromised can cause serious effects to agents. Similarly, fake information (incorrect facts, misleading arguments), made-on-purpose contamination with the only goal to pollute the memory and influence future behavior, and benign-looking triggers (information from an agent that appears harmless but that can activate poisoned memory entries) may cause undesired consequences. Research needs to focus on this type of interaction-based attacks.

Nevertheless, these types of interactions and attacks are quite different from those described above by Chen et al.~\cite{ref:Chen.et.al.2024:NeurIPS}. Successful attacks will not be a set of facts (or key-value pairs) added to a static memory, but a set of successive interactions (maybe long term ones to avoid detection) added to a dynamic memory. I.e., there is no static ${\cal{D}}_{clean}$ memory to consider. This is a much complex scenario, more particularly, because, as Rando et al.~\cite{ref:Rando.Zhang.Carlini.Tramer.2025} describe, {\it adversarial ML problems are getting harder to solve and to evaluate}. 

\section{Conclusions}
Memory poisoning is being considered a threat for the proper development of multi-agent systems. In this paper we discuss mitigation strategies based on security technologies. More particularly, we discuss threats and mitigation strategies for three memory systems: semantic, episodic, and short-term memories. 

We claim that agents need to be built using secure memory, interactions need to be secure, and information needs to be signed and acquired only from trusted agents. We propose to use secure episodic and short-term memories. Private knowledge retrieval and private inference should be used to obtain knowledge from untrusted agents. We have described an approach for performing private inference using private information retrieval, and we have implemented a Python-based proof-of-concept for a Prolog-like (Horn clauses) inference engine that employs private knowledge retrieval by means of accessing two knowledge bases. We have also implemented a lighter solution with fewer privacy guaranties, which relies on a single knowledge base and is based on k-anonymity.

Provenance structures need to be implemented for short-term and episodic memories. This provides information about the items in the memory (who, what, and when the information was provided) that can be used in the decision making process itself, but also can help in the event of attacks. In this way, provenance provides information that can be analyzed not only to detect attacks but also used for digital forensics if attacks are successful. 

Finally, reputation and trust mechanisms need to be in place in multi-agent systems to leverage information from other agents in the community and build trustworthy systems. 

Both provenance structures and reputation mechanisms play a key role in building episodic memory. They help secure short-term memory information and support secure memory updating algorithms. 

From our perspective, two key research questions that need to be studied in more detail are

\begin{itemize}
\item Secure episodic memory update algorithms. Current algorithms ignore completely security issues, and they focus on e.g. recall of memory items. 
\item Memory manipulation through agent interactions. Poisoning short-term memory by means of purposeful interactions can cause short- and long-term damage. Nevertheless, it is unclear how these interactions can affect future behavior, except in unrealistic cases. 
\end{itemize}

\section*{Acknowledgements}
This work was partially supported by the Wallenberg AI, Autonomous Systems and Software Program (WASP) funded by the Knut and Alice Wallenberg Foundation. Support by Swedish Research Council (project VR 2023-05531) is also acknowledged. 	
It was also partially supported by 
project PID2024-156636NB-C21 (MATSE) funded by MCIN/AEI/10.13039/501100011033/ FEDER, UE.



\end{document}